\scrollmode
\documentstyle[draft]{mn}

\input epsf

\title[An evolutionary scenario..]{An evolutionary scenario for the U Scorpii}

\author[Ene Ergma et al.]{Ene Ergma$\rm ^{1}$, Jelena Ger\v{s}kevit\v{s}$\rm 
^{1,2}$ and Marek J. Sarna$\rm ^{2}$ \\
$\rm  ^1~$ Physics Department, Tartu University, \"Ulikooli 18, 50510
       Tartu, Estonia \\
       e--mail: ene$@$physic.ut.ee; jelen\_a@physic.ut.ee \\
$\rm  ^2~$ N. Copernicus Astronomical Center,
       Polish Academy of Sciences,
       ul. Bartycka 18, 00--716 Warsaw, Poland \\
       e--mail: sarna$@$camk.edu.pl; jelena@camk.edu.pl \\}

\date{Received; accepted}

\begin{document}

\maketitle

\begin{abstract}
We perform evolutionary calculations of binary stars  to find progenitors of 
systems with parameters similar to the recurrent novae U Sco.  We show 
that a U Sco type--system may be formed starting with an initial binary 
system which has a low--mass carbon-oxygen white dwarf as an accretor. 
Since the evolutionary stage of 
the secondary is not well known, we calculate sequences with 
hydrogen rich and helium rich secondaries. The evolution of the binary    
may be devided into several observable stages as: classical nova, 
supersoft X--ray source with 
hydrogen stable burning and strong  wind phases, ending up with the formation 
of a massive  white dwarf near the Chandrasekhar mass limit. 
We follow the chemical evolution of the secondary as well as of the matter 
lost from 
the system, and we show that observed $^{12}$C/$^{13}$C and 
N/C ratios may give some information about the nature of the binary.   
\end{abstract}


\section{Introduction}

Recurrent novae are a small class of objects which bear many similarities 
to other cataclysmic variable systems. They experience recurrent outbursts 
at intervals of 20--80 yrs.

Webbink et al. (1987) lengthily discussed the nature of the recurrent novae, 
and they concluded that according to outburst mechanisms there are 
two subclasses of these systems: (a) powered by thermonuclear 
runaway on the surface of the white dwarf (e.g. U Sco), 
and (b) powered by the transfer of a burst of matter from the red giant 
to the main--sequence companion. 

U Sco is one of the best observed recurrent novae. Historically, its
outbursts were observed in 1863, 1906, 1936, 1979, 1987 and in 1999.
Determinations of the system visual luminosity at maximum and minimum,
indicate a range $\Delta m_V \sim $9. We also note that the mean recurrent
interval implied by known outbursts is $P_{rec} = 23 ~$yrs. 
Schaefer (1990) and Schaefer \& Ringwald (1995) observed eclipses of U Sco in 
the quiescent phase, and determined the orbital period $P_{orb}$=1.23056 d.

Ejecta abundances have been estimated (from 1979 outburst) from optical 
and UV studies by Williams et al. (1981) and Barlow et al. (1981). They 
derived extremely helium rich ejecta He/H$\sim$ 2 (by number), while the 
CNO abundance was solar with an enhanced N/C ratio.
From the analysis of the 1999 outburst Anupama \& Dewangan (2000) obtained an average 
helium abundance of He/H$\sim$0.4$\pm$0.06. The estimated mass of the 
ejected shell for 1979 and 1999 outbursts is $\rm \sim 10^{-7} ~M_\odot$ 
(Williams et al. 1981; Anupama \& Dewangan 2000). Spectroscopically, U Sco 
shows very high ejection velocities of (7.5--11)$\rm \times 10^3 ~km ~s^{-1}$ 
(Williams et al. 1981; Munari et al. 2000). Latest determinations of the 
spectral type of the secondary indicate a K2 subgiant (Anupama \& Dewangan 
2000; Kahabka et al. 1999). Following to  Kahabka et al. (1999) the distance 
to U Sco is about 14 kpc.

According to Kato model (1996), supersoft X--ray emission is predicted 
to be observed about 10--60 days after the optical outburst. BeppoSAX 
detected  supersoft X-ray emission from U Sco in range 0.2--20 keV just 
19--20 days after the peak of its optical outburst in February 1999 
(Kahabka et al.  1999). The fact that U Sco was detected as a supersoft 
X--ray source (SSS) is consistent with steady hydrogen burning on the 
surface of its white dwarf component.

In this paper we construct several evolutionary sequences which may lead 
to formation of systems like U Sco. In Section 2 the evolutionary code is 
briefly described. In Section 3 we discuss the effect of mass transfer on 
binary evolution. Section 4 contains the results of the calculations.
A general discussion and conclusion follow.

\section{The evolutionary code} 

The models of secondary stars filling their Roche lobes were computed
using a standard stellar evolution code based on the Henyey--type code
of Paczy\'nski (1970), which has been adapted to low--mass stars (Marks \& 
Sarna 1998, hereafter MS98).

Our nuclear reaction network is based on that of Kudryashov \& Ergma (1980), 
who included the reactions of the CNO tri--cycle in their calculations of
hydrogen and helium burning in the envelope of an accreting neutron
star. We have included the reactions
of the proton--proton (PP) chain. Hence we are able to follow the
evolution of the elements: $^{1}$H, $^{3}$He, $^{4}$He, $^{7}$Be,
$^{12}$C, $^{13}$C, $^{13}$N, $^{14}$N, $^{15}$N, $^{14}$O, $^{15}$O,
$^{16}$O, $^{17}$O and $^{17}$F. We assume that the abundances of
$^{18}$O and $^{20}$Ne stay constant throughout the evolution. We use
the reaction rates of: Fowler, Caughlan \& Zimmerman (1967, 1975), Harris at 
al. (1983), Caughlan et al. (1985), Caughlan \& Fowler (1988), Bahcall \& 
Ulrich (1988), Bahcall \& Pinsonneault (1992), Bahcall, Pinsonneault \& 
Wesserburg (1995) and Pols et al. (1995). We use the
Eggleton (1983) formula to calculate the size of the secondary's
Roche lobe.

For radiative transport, we use the opacity tables of
Iglesias \& Rogers (1996). Where they are 
incomplete, we fill the gaps using opacity tables of 
Huebner et al. (1977). For temperatures lower than 6000 K we use the 
opacities given by Alexander \& Ferguston (1994) and Alexander  
(private communication). For a more detailed description of the code see 
MS98.

\section{The effects of mass transfer}

While calculating evolutionary models of binary stars, we must take into
account mass transfer and associated physical mechanisms which lead to
mass and angular momentum loss. 
Apart from angular momentum loss due to gravitational wave radiation
and magnetic braking, there is additional loss due to
the non--conservative nature of semidetached evolution, such as novae
outbursts and strong optically thin/thick wind from white dwarf occuring 
during stable hydrogen shell burning phase. 

We can express the change in the total orbital angular momentum ($J$)
of a binary system as

\begin{equation}
\frac{\dot{J}}{J} = \left. \frac{\dot{J}}{J} \right|_{\rm GWR} +
\left. \frac{\dot{J}}{J} \right|_{\rm MSW} + \left. \frac{\dot{J}}{J}
\right|_{\rm NOAML} + \left. \frac{\dot{J}}{J} \right|_{\rm FAML}
+ \left. \frac{\dot{J}}{J} \right|_{\rm FWIND}
\end{equation}

where the terms on the right hand side are due to: gravitational wave
radiation, magnetic stellar wind braking, novae outbursts angular
momentum loss (which describe the loss of angular momentum from the
system due to non--conservative evolution), frictional angular
momentum loss (which occurs during a novae outburst as the secondary
orbits within the expanding novae shell and during strong optically thin/thick 
wind) and optically thin/thick wind from white dwarf (during stable 
hydrogen shell burning). The role of the first two terms have been discussed 
in many papers (see, for example, review paper by Verbunt 1993). 
In our paper we discuss the last three terms.

\subsubsection{Angular momentum loss associated with novae outbursts}

To take account of the angular momentum loss that accompanies mass
loss due to novae outbursts that occur on the surface of the white
dwarf during the semidetached phase, we use a formula based on that
used to calculate angular momentum loss via a stellar wind
(Paczy\'nski (1967); Zi\'o\l kowski (1985) and De Greve (1993)),

%--------------------------------------------------------------------%
\begin{equation}
\left. \frac{\dot{J}}{J} \right|_{NOAML} = f_{1} \: f_{2} \:
\frac{M_{wd} \: {\dot{M}}_{sg}}{M_{sg} \: M_{tot}}, \:\:\:\:\: {\rm where}
\:\:\:\:\: \dot{M} = f_{1} \: {\dot{M}}_{sg}.  
\end{equation}
%--------------------------------------------------------------------%

$f_{1}$ is the ratio of the mass ejected by the white dwarf to
that accreted by the white dwarf; $f_{2}$ is defined as the
effectiveness of angular momentum loss during mass transfer
(Sarna \& De Greve 1994, 1996); $ M_{wd} $ and $ M_{sg} $ denote mass  
of white dwarf primary and subgiant secondary, respectively; $M_{tot} $ 
(=$\rm M_{wd} + M_{sg} $) is the total
mass of the system; $\dot{M}$ and ${\dot{M}}_{sg}$ are, respectively, 
the rate of mass loss from the system and the rate of mass loss from
the secondary ($-$ ${\dot{M}}_{sg}$ is equivalent to the mass
transfer rate). We take $f_{2} = 1.0$ which is typical
for a stellar wind. Recently, Livio \& Pringle (1998) proposed a model 
in which the accreted angular momentum is removed from the system 
during novae outbursts, 
which agrees with our earlier suggestions (Marks, Sarna \& Prialnik 1997, 
MS98). Equation (2) is in quantitative agreement with estimations made 
by Livio \& Pringle (1998).

\subsection{Frictional angular momentum loss}

During nova outburst or optically thin/thick wind phase the 
secondary star effectively orbits within
the expanding nova shell or dense wind matter. Due to the frictional 
deposition of orbital
energy into expanding nova shell or strong wind, the separation of the 
components will be decreased.

Livio, Govarie \& Ritter (1991) estimated the change in the orbital angular 
momentum brought about by frictional angular momentum loss over a
complete novae cycle,

%--------------------------------------------------------------------%
\begin{equation}
\left. \frac{\dot{J}}{J} \right|_{\rm FAML} = \frac{1}{4}
\left( 1 + q \right) \: \frac{\left( 1 + U^{2}
\right)^{\frac{1}{2}}}{U} 
\left( \frac{R_{sg}}{a}\right)^{2} \: \frac{{\dot{M}}_{\rm wind}}{M_{sg}} , 
\end{equation}
%--------------------------------------------------------------------%

where $U$ is the ratio of the expansion velocity of the envelope at
the position of the secondary to the orbital velocity of the secondary
in the primary's frame of reference ($v_{\rm exp}$/$v_{\rm orb}$),
($q=M_{wd}/M_{sg}$) is the mass ratio  and
${\dot{M}}_{\rm wind}$ is the rate of mass flow past the secondary
Warner (1995). For strong winds from white dwarf during 
stable hydrogen shell burning Hachisu, Kato \& Nomoto (1996) 
estimated $v_{\rm exp} / v_{\rm orb} \sim 10$. Since we have no
information concerning the expansion velocity of the ejecta from  
theoretical novae models of Prialnik \& Kovetz (1995) and Kovetz \& Prialnik
(1997) (hereafter PK95 and KP97), we do not include the effect of frictional
angular momentum loss during nova phase. We include this effect during
strong wind phase.

\subsection{Optically thin/thick wind angular momentum loss}

If we consider a carbon--oxygen (C--O) white dwarf accreting matter from a 
companion 
with solar composition, there exists a critical accretion rate above 
which the excess material is blown off by strong wind. 
Hachisu et al. (1996) show that because wind velocity is about
10 times higher than orbital velocity, the wind has the same 
specific angular momentum as that of the white dwarf, which is estimated 
as

%-----------------------------------------------------------------------%
\begin{equation}
\left. \frac{\dot{J}}{J} \right|_{\rm FWIND} = {q \over {1 +q}} 
{{{\dot M}_{wind}} \over M_{sg}} . 
\end{equation}
%-----------------------------------------------------------------------%

As argued by MS98, for typical
binary parameters, angular momentum loss due to nova outbursts and magnetic stellar
wind braking are the dominant angular momentum loss mechanisms, with
gravitational wave radiation two orders of magnitude less effective
(when the orbital period decreases
sufficiently, gravitational wave radiation will be more effective).
Frictional angular momentum loss is the least effective at eight
orders of magnitude less than novae outbursts angular momentum loss and
magnetic stellar wind braking. However this mechanism will be very effective 
during strong optically thin/thick wind (SSS phase). 
Note also, that during the wind phase the wind carries off the specific angular momentum of 
the white dwarf, which stabilizes the mass transfer (Hachisu et al. 1996;
Li \& van den Heuvel  1997). 

\subsection{Accretion of material ejected during novae outbursts}

We employ the code developed by MS98, which utilises the 
results of the theoretical novae calculations made by PK95 and KP97. 
By interpolation from the data sets of PK95 and KP97, at each
time--step, we use the white dwarf mass and the mass
transfer rate to determine the novae characteristics: $f_1$, amplitude 
of the outburst, recurrence period, chemical composition of the ejected 
material. 

To calculate the re--accretion by the secondary of material ejected
during novae outbursts, we assume that the mass of the material
re--accreted ($M_{\rm re-acc}$) is proportional to the mass of the
material ejected by the white dwarf such that,

%--------------------------------------------------------------------%
\begin{equation}
M_{\rm re-acc} = \left( \frac{{R_{sg}}^{2}}{4 a^{2}} \right) M_{\rm ej}
\:\:\:\:\:\: ,
\end{equation}
%--------------------------------------------------------------------%

where $M_{ej}$ is the amount of matter ejected in the nova outburst, 
$R_{sg}$ and $a$ are the radius of the secondary star and the
separation of the system, respectively. The constant of proportionality
is the ratio of the cross--sectional area of the secondary star to the
area of a sphere at radius $a$ from the white dwarf. We base this
formula on the assumption that novae ejections are spherically
symmetric and instantaneous. 

\subsection{Accretion of material from strong wind}

We define two critical mass accretion rate onto white dwarf. First (Warner 
1995):

%---------------------------------------------------------------------%
\begin{equation}
{\dot M}_{cr,1} = 2.3 \times (M_{wd} - 0.19)^{3/2} ,
\end{equation}
%---------------------------------------------------------------------%

describing critical accretion rate above which hydrogen rich material is
burning in a stable shell; and below which novae outbursts occur.
Second (Nomoto, Nariai \& Sugimoto 1979; Hachisu et al. 1996):

%----------------------------------------------------------------------%
\begin{equation}
{\dot M}_{cr,2} = 9.0 \times (M_{wd} - 0.5) ,
\end{equation}
%----------------------------------------------------------------------%

describing the critical accretion rate above which strong wind solution by 
Hachisu et al. (1996) is working. It allows burning of the hydrogen 
into helium at a rate close to ${\dot M}_{cr,2} $, with the excess material 
being blown off by the wind at the rate:

%----------------------------------------------------------------------%
\begin{equation}
{\dot M}_{wind} = {\dot M}_{sg} - {\dot M}_{cr,2} .
\end{equation}
%----------------------------------------------------------------------%

To calculate re-accretion of material from the wind by the secondary, we 
assume that, similarly to eq. 5, the mass of re-accreted material is 
proportional to the mass of material loss by the white dwarf due to 
strong wind (${\dot M}_{wind} $).

\section{Results of calculations}

Since the evolutionary stage of the secondary (hydrogen or helium rich) is 
not observationally determined (more details in section 5), we computed four 
different sequences with hydrogen and helium rich secondaries.  Initial 
parameters of the sequences are presented in Table 1.

\begin{tabular}{lccccc}
\multicolumn{6}{l}{Table 1 ~~Initial parameters of computed sequences} \\
\hline
model  &  $\rm M_{sg} $ & $\rm M_{wd}$ & $\rm P_{i}$(RLOF) & X & Z \\
 & [$\rm M_\odot$] & [$\rm M_\odot$] & [d] & & \\
\hline
A        & 1.4 & 0.70 & 1.34 & 0.7 & 0.02 \\
B        & 1.7 & 0.85 & 1.61 & 0.7 & 0.02 \\
C        & 1.4 & 0.70 & 1.26 & 0.5 & 0.02 \\
D        & 1.7 & 0.85 & 1.24 & 0.5 & 0.02 \\
\hline
\end{tabular}

The sequences C and D present a helium rich SSS channel 
proposed by Hachisu et al. (1999). In Table 2 we present results 
for computed sequences. The orbital parameters are given at the 
moment when orbital period is equal to 1.23 d. The effective temperature and 
luminosity are for subgiant star.

According to our calculations, all models go through the short time 
first nova 
phase (n1). We have used the grid of models calculated by PK95, KP97 and 
their classification scheme to relate theoretical models to observations of 
nova outburst (for more details see MS98). After that, systems enter 
into the first stable hydrogen burning phase (s1), which is followed by the 
wind phase (w). The sequences B, C and D have wind phase and second stable 
hydrogen burning phase (s2). The sequence A does not exhibit wind phase, 
and after stable burning phase this system evolves  into
second nova phase (n2). After second stable hydrogen burning phase 
sequences B and C evolve through recurrent novae phase. In Table 3 we give  
the duration of each phase. The duration of the SSS stage 
(stable hydrogen burning and wind phases) is 
ranging from several hundred thousand to several million years. 
During the second nova phase, sequence A shows behaviour characteristic for slow 
novae. Sequence D avoids second novae phase, and for this system  the 
white dwarf mass will exceed Chandrasekhar limit during stable 
hydrogen burning, and a supernova explosion may occur. 

\begin{figure}
\epsfverbosetrue
\begin{center}
\leavevmode
\epsfxsize=7cm
\epsfbox{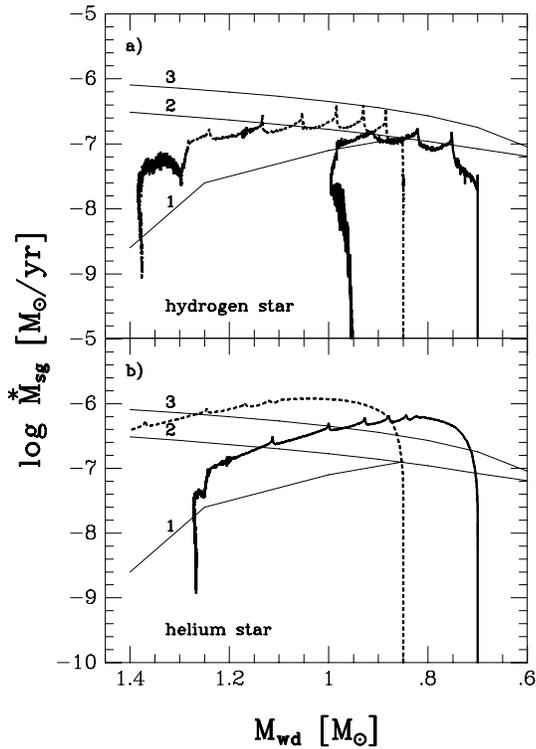}
%\plottwo{fig1.ps}{fig2.ps}
\end{center}
\caption{The evolution of the mass transfer rate versus white 
dwarf mass: (a) for hydrogen rich star as donor (sequences A, B in Table 1), 
and (b) for helium rich star as donor (sequences C, D in Table 1). 
The stable hydrogen burning and wind solutions are shown. The lines marked 
1, 2 and 3 shows lower boundaries for recurrent novae, stable hydrogen 
burning and strong wind regions, respectively. The region restricted by 
lines 1 and 2 shows the recurrent nova phase.} 
\end{figure}

U Sco progenitors 
can be found with the help of evolutionary sequences 
B (hydrogen rich) and C (helium rich).  After stable hydrogen burning 
stage both sequences enter into recurrent nova phase (Fig.1).
In Fig. 2 the evolution of the mass accretion rate versus orbital period 
is shown for two sequences B and C. From Fig. 2 we see that both systems 
evolve through the orbital period $P_{orb}$ =1.23 d twice. During the first 
crossing of $P_{orb}=1.23 d $ line the luminosity of the secondary is too high 
(log L/$\rm L_\odot$$\sim$0.9 and 1.35 for sequences B and C, respectively) 
and it does not fit the observed absolute magnitude  $M_{V}$=+3.8 of U Sco.

\begin{figure}
\epsfverbosetrue
\begin{center}
\leavevmode
\epsfxsize=7cm
\epsfbox{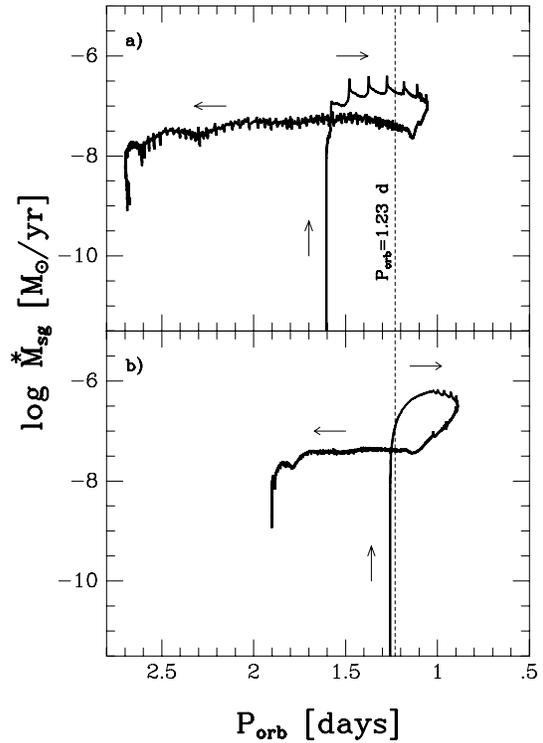}
%\plottwo{fig1.ps}{fig2.ps}
\end{center}
\caption{The evolution of the mass transfer rate versus the orbital 
period of the system: (a) for hydrogen rich star as donor (sequence B in 
Table 1, 2), and (b) for helium rich star as donor (sequence C in Table 1, 2). 
The arrows show the directions of evolution. The vertical (thin dashed) line 
marks the position of the U Sco orbital period.} 
\end{figure}

\begin{figure}
\epsfverbosetrue
\begin{center}
\leavevmode
\epsfxsize=7cm
\epsfbox{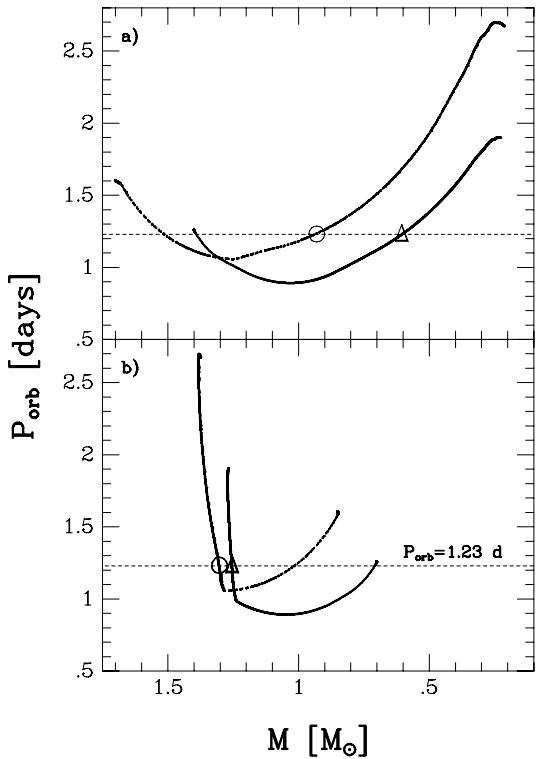}
%\plottwo{fig1.ps}{fig2.ps}
\end{center}
\caption{The evolution of the orbital period of the system as a function
(a) of the subgiant mass $M_{sg}$: solid line -- sequence C, dashed line -- 
sequence B, and (b) of the white dwarf mass $M_{wd}$: solid line -- sequence 
C, dashed line -- sequence B. Horizontal thin dashed lines
mark orbital period for U Sco, open circle mark positions of subgiant and
white dwarf masses for sequence B, open triangle for sequence C. 
For upper panel evolution is going from left to right, for lower panel 
from right to left.} 
\end{figure}

For the same two evolutionary sequences, in Fig. 3 we present evolution 
of the orbital period $P_{orb}$ versus mass 
of subgiant $M_{sg}$ (Fig. 3a) and mass of the white dwarf $M_{wd}$ (Fig.
3b). From grid of models (PK95 and KP97) we find $\it A$=7.6 mag, and 
$\rm P_{rec} $= 23 and 54 yrs for sequences B and C, respectively. 
Therefore, we can conclude that sequence B gives the best fit to observing 
parameters of U Sco. 

\begin{table}
\begin{center}
\begin{tabular}{lccccc}
\multicolumn{6}{l}{Table 2 ~~Results for computed sequences} \\
\hline
model  &  $\rm M_{sg} $ & $\rm M_{wd}$ & $\rm \log T_{eff} $
 & log $\rm L/L\odot $ & $\dot{M}_{sg}$ \\
 & [$\rm M_\odot $] & [$\rm M_\odot $] & [K] & & [$\rm M_\odot ~yr^{-1} $] \\
\hline
A  &0.545   &0.983   &3.669  &0.153  &2.02$\times 10^{-8}$ \\
B  &0.936   &1.304   &3.693  &0.416  &3.58$\times 10^{-8}$ \\
C  &0.603   &1.255   &3.721  &0.388  &4.09$\times 10^{-8}$ \\
D  &1.696   &0.852   &3.957  &1.707  &1.14$\times 10^{-8}$ \\ 
\hline
\end{tabular}
\end{center}
\end{table}

\begin{table}
\begin{center}
\begin{tabular}{lccccc}
\multicolumn{6}{l}{Table 3 ~~Time--scales for evolutionary phases} \\
\hline
model & $\rm \Delta t^1_{novae}$ & $\rm \Delta t^1_{stat} $ & $\rm \Delta
t_{wind} $ & $\rm \Delta t^2_{stat} $ & $\rm \Delta t^2_{novae} $ \\
\multicolumn{6}{c}{[$\rm \log (\Delta t/ yr) $]} \\
\hline
A   &4.00     &6.57   &--     &--    &8.17    \\ 
B   &5.97     &5.53   &3.70   &6.40  &7.05    \\
C   &4.76     &5.41   &5.39   &6.17  &7.30    \\
D   &4.89     &5.02   &5.60   &5.58  &--      \\  
\hline
\end{tabular}
\end{center}
\end{table}

\subsection{Chemical composition of the subgiant and ejected matter}

Our program is able to follow in detail the evolution of the chemical composition 
of the subgiant and the ejected matter. In Table 4 we show isotopic 
composition in the envelope of the subgiant and the ejected matter for 
two sequences B and C, for a moment when the binary system has orbital period 
$P_{orb}$=1.23 d.

\begin{table*}
\begin{center}
\begin{tabular}{lcrrrcrcr}
\multicolumn{9}{l}{Table 4 ~~Chemical compositions for sequence B and C} \\
\hline
model & $^{12}$C & $^{13}$C & $^{14}$N & $^{15}$N  &
 $^{16} $O & $^{17}$O & $^{12}$C/$^{13}$C  &  $\rm N/C $ \\
& [$\rm \times 10^{-3} $] & [$\rm \times 10^{-4} $] & [$\rm \times 10^{-3} $]
& [$\rm \times 10^{-7} $] & [$\rm \times 10^{-3} $] & [$\rm \times 10^{-5} $]
& & \\
\hline
B subgiant & 1.41 & 3.05  & 3.16  & 5.64  & 9.77 & 0.58  & 4.6 & 1.8  \\
B ejecta   & 2.81 & 11.51 & 42.06 & 251.60 & 1.77 & 200.00 & 2.4 & 10.6 \\   
C subgiant & 0.41 & 1.40  & 4.53  & 2.71  & 9.79 & 1.17   & 2.9 & 8.2  \\
C ejecta   & 1.94 & 6.97  & 37.10 & 83.78 & 4.94 & 68.48  & 2.8 & 14.0 \\
\hline
\end{tabular}
\end{center}
\end{table*}

If we compare the helium rich model C with the hydrogen rich model 
B we can see that N/C ratio in the envelope of the subgiant is higher for 
helium rich model than hydrogen rich one, but the  
isotopic ratio $^{12}$C/$^{13}$C is higher for the hydrogen rich model. 
In the ejected matter both ratios are similar. 

Our theoretical calculations show that He/H ratio for the matter lost 
from the system changes during evolution from 0.56 to 1.26. 
For evolutionary sequences B and C and for the orbital period of 
U Sco, this ratio is about 0.7. Our He/H determination is well inside 
observational determination which vary between 0.4 and 2 
(Anupama \& Dewangan 2000, Williams et al. 1981). 

\begin{figure}
\epsfverbosetrue
\begin{center}
\leavevmode
\epsfxsize=7cm
\epsfbox{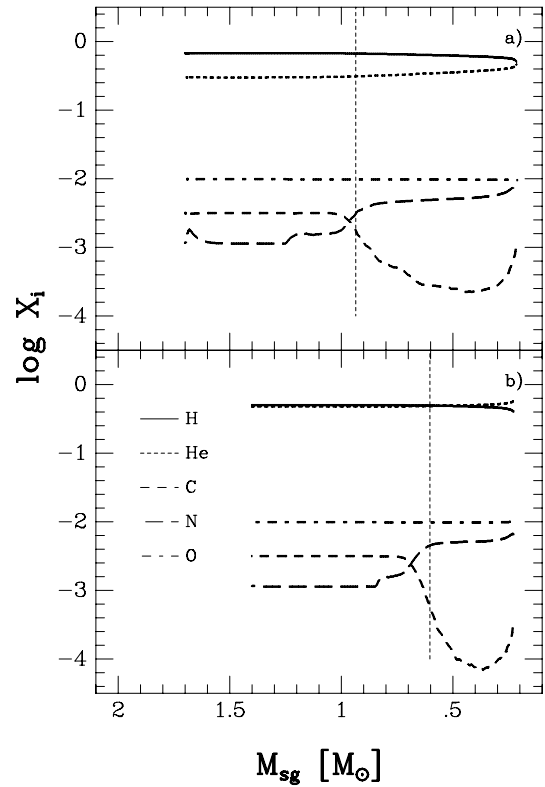}
%\plottwo{fig1.ps}{fig2.ps}
\end{center}
\caption{The evolution of the red giant surface abundances of H, He, C, N and O 
as a function of the subgiant mass $M_{sg}$: upper panel -- sequence B, 
lower panel -- sequence C. Vertical thin dashed lines mark the position 
of subgiant mass for sequences B ($\rm M_{sg}=0.926 ~M_\odot $) and C 
($\rm M_{sg}=0.603 ~M_\odot $) respectively (see Table 2 for more details). 
} 
\end{figure}

Figs. 4 and 5 show the evolution of the abundances of helium, hydrogen, 
carbon, nitrogen and oxygen of the subgiant envelope and ejected matter. 
The vertical thin dashed lines show the place where the orbital period is 
equal to 1.23 d. In Fig.5 we also identify the phases of the binary
system evolution: n1 -- first short time nova episode, s1 -- first stable
hydrogen burning phase, w -- strong wind phase, s2 -- second stable hydrogen
burning phase and n2 -- second nova phase.

\begin{figure}
\epsfverbosetrue
\begin{center}
\leavevmode
\epsfxsize=7cm
\epsfbox{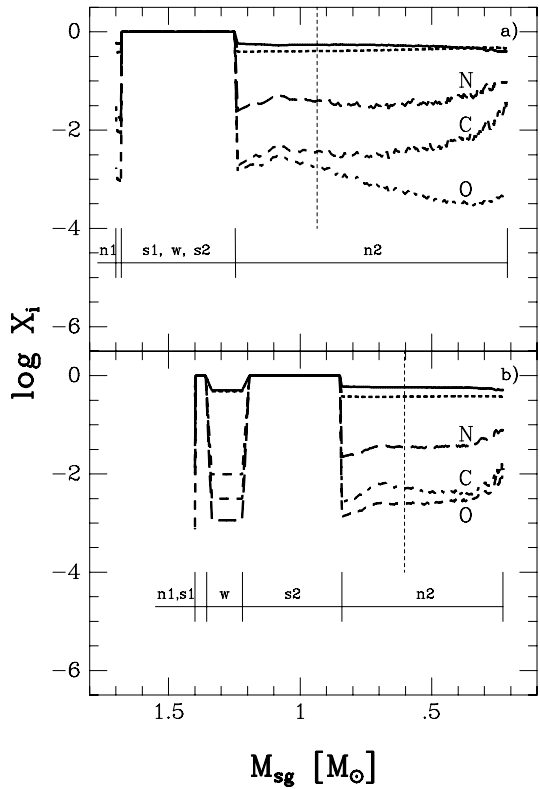}
%\plottwo{fig1.ps}{fig2.ps}
\end{center}
\caption{The evolution of the ejected matter abundances of H, He, C, N and O 
as a function of the subgiant mass $M_{sg}$: upper panel -- sequence B, 
lower panel -- sequence C. For more explanation see text.} 
\end{figure}

Chemical composition and isotopic analysis may give more information 
about the evolutionary stage of U Sco. Possible observational tests 
are discused in detail by MS98. Unfortunately, the subgiant
component in U Sco is too faint for infrared spectroscopic observations of 
CO bands in order to determine the $^{12}$C/$^{13}$C ratio.
However, we think that blue domain spectra of U Sco could show some
absorption structure in the region of 4216\AA~ in the CN sequence like one
observed for DQ Her (Chanan, Nelson \& Margon 1978, Schneider \& Greenstein 
1979, Willimas 1983). Analysis of this region of the spectra is more 
complicated because the structure of the CH and CN violet system can be affected 
by some absorption features from the disc. However, we suggest that since the matter 
in the accretion disc reflects the chemical composition of the subgiant star,
analysis of the disc will give us the information we need if we use
observations made during the quiescent phase. The chemical analysis of the
expanding envelope (Anupama \& Dewangan 2000) also will give useful 
information allowing comparison with theoretical models (see Table 4).

\section{Discussion}

Hachisu et al. (1999) proposed a new evolutionary path to SNe Ia, in which 
the companion star is helium--rich. In their model, typical 
orbital parameters of SNe Ia progenitors are: 
$M_{wd}$ =1.37$\rm ~M_\odot$, 
$M_{sg}$$\sim$ 1.3$\rm ~M\odot$, and $\dot{M}_{sg}\sim$ 2$\times 
10^{-7}$$\rm ~M_\odot ~yr^{-1}$. Based on light--curve analysis, 
Hachisu et al. (2000) also constructed a detailed theoretical model for 
U Sco. They found that the best fit 
parameters are: $M_{wd}$$\sim$ 1.37$\rm ~M_\odot$, 
and $M_{sg} \sim$ 1.5$\rm ~M_\odot$ (a range from 0.8 to 2.0$\rm ~M_\odot$ is 
acceptable).

However, the helium enriched model poses a serious problem. Truran et al. 
(1988) 
discussed the composition dependence of thermonuclear runaway models for 
the recurrent novae of U Sco--type. They showed that for $\rm M_{wd} = 1.38 ~M_\odot$, 
$\dot{M}_{sg}$
=$\rm 1.5 \times 10^{-8} ~M_\odot ~yr^{-1} $, L=0.1$\rm ~L_\odot$ optically bright 
outbursts are obtained only for matter with He/H$<$1--2. 
Above results are consistent with our sequences B and C where He/H are 
equal 0.48 and 0.96, respectively.
 
There is an alternative explanation of helium enrichment as observed in
U Sco which may also occur due to helium enriched winds from the white dwarf 
(Prialnik \& Livio 1995).

According to Kato (1996), the supersoft 
component in UV spectrum is predicted to be observable 
about 10 days after the outburst. 
For hydrogen--rich model (He/H=0.1) the supersoft X--ray component is expected to 
rise till $\sim$ 50 days after the outburst to a maximum luminosity of 
$\sim$ 3$\times 10^{36}$erg$s^{-1}$ $(d/kpc)^2$. For helium--rich model 
(He/H = 2), the maximum luminosity is reached about 20 days after the optical 
outburst.  Unfortunately, Kahabka et al. (1999) could 
follow the X--ray outburst of U Sco for only  
19--20 days after outburst. According to Kato (1996), the X--ray luminosity 
behaviour depends on the chemical composition. Therefore, the evolution 
of the X--ray luminosity may give the important evidence about the chemical 
composition of the accreted matter.

\section{Conclusion}

We calculated several evolutionary sequences to reproduce 
orbital and physical parameters of the recurrent novae of 
U Sco--type. We showed that U Sco systems  
possibly form from binaries with a low--mass C--O white dwarf as accretor. 
Such a system evolves through several observable stages: recurrent nova, SSS  
with stable hydrogen burning and SSS with strong 
wind phases. In final phase of evolution a massive
white dwarf near the Chandrasekar limit is formed. 
We propose that the evolutionary 
sequence B is able to produce binary system with parameters similar to U Sco. 
Our best fitting model has initial parameters: $M_{sg,i} = 1.7 ~M_\odot $, 
$M_{wd,i} = 0.85 ~M_\odot $ and $P_i (RLOF) = 1.61 $d. Based on evolutionary
model we constracted a detailed theoretical model for U Sco. We found that
the best fit parameters are: $M_{sg} = 0.94 ~M_\odot $, $M_{wd} = 1.31
~M_\odot $, $\log L_{sg}/L_\odot = 0.42 $, $\dot M_{sg} = 3.58 \times 10^{-8}
M_\odot ~yr^{-1} $ for $P_{orb} = 1.23056 ~d$. 

\section*{\sc Acknowledgments}

This work is partly supported through grant 2--P03D--005--16 of the 
Polish National Committee for Scientific Research. JG and EE 
acknowledge support through Estonian SF grant 4338. EE  acknowledges warm
hospitality of the Astronomical Institute ``Anton Pannekoek'' where part of 
this work has been conducted. While in Netherlands, EE was supported by 
NWO Spinoza grant 08--0 to E. P. J. van den Heuvel.

\end{document}